\documentclass[pdftex, aps, prd, a4paper, superscriptaddress, nofootinbib,
               10pt, preprintnumbers, longbibliography]{revtex4-1}
\pdfoutput=1
\usepackage[utf8]{inputenc}
\usepackage{amsmath, amsfonts, amssymb, xcolor,cancel,soul}
\usepackage{textcomp}
\usepackage[colorlinks=true, linkcolor=blue, filecolor=blue, urlcolor=blue,
            citecolor=blue,  plainpages=false]{hyperref}

\begin{document}
\title{A lattice QCD perspective on weak decays of \texorpdfstring{$b$ and $c$}{b and c}
       quarks \texorpdfstring{\\}{ } Snowmass 2022 White Paper}

\author{Peter A.~Boyle}
\affiliation{Physics Department, Brookhaven National Laboratory, Upton, NY 11973, USA}
\affiliation{Higgs Centre for Theoretical Physics, The University of Edinburgh, EH9 3FD, UK}

\author{Bipasha Chakraborty}
\affiliation{DAMTP, Centre for Mathematical Sciences, University of Cambridge, Wilberforce Road, Cambridge, CB3 0WA, UK}

\author{Christine T.~H.~Davies}
\affiliation{SUPA, School of Physics and Astronomy, University of Glasgow, Glasgow, G12 8QQ, UK}

\author{Thomas DeGrand}
\affiliation{Department of Physics, University of Colorado, Boulder, CO 80309 USA}

\author{Carleton DeTar}
\affiliation{Department of Physics and Astronomy, University of Utah, Salt Lake City, UT 84112, USA}

\author{\\Luigi Del Debbio}
\affiliation{Higgs Centre for Theoretical Physics, The University of Edinburgh, EH9 3FD, UK}

\author{Aida X.\ El-Khadra}
\affiliation{Department of Physics, University of Illinois at Urbana-Champaign, Urbana, IL 61801, USA}

\author{Felix Erben}
\affiliation{Higgs Centre for Theoretical Physics, The University of Edinburgh, EH9 3FD, UK}

\author{Jonathan M.~Flynn}
\affiliation{Physics and Astronomy, University of Southampton, Southampton SO17 1BJ, UK}

\author{Elvira G\'amiz}
\affiliation{Theoretical Physics Department, University of Granada, E-18071 Granada, Spain}

\author{\\Davide Giusti}
\affiliation{Universit\"at Regensburg, Fakult\"at f\"ur Physik, 93040 Regensburg, Germany}

\author{Steven Gottlieb}
\affiliation{Department of Physics, Indiana University, Bloomington, IN 47405, USA}

\author{Maxwell T.~Hansen}
\affiliation{Higgs Centre for Theoretical Physics, The University of Edinburgh, EH9 3FD, UK}

\author{Jochen Heitger}
\affiliation{Westf\"alische Wilhelms-Universit\"at M\"unster, Institut f\"ur Theoretische Physik,
48149 M\"unster, Germany}

\author{Ryan Hill}
\affiliation{Higgs Centre for Theoretical Physics, The University of Edinburgh, EH9 3FD, UK}

\author{\\William I.~Jay}
\affiliation{Center for Theoretical Physics, Massachusetts Institute of Technology, Cambridge, MA 02139, USA}

\author{Andreas J\"uttner}
\affiliation{Physics and Astronomy, University of Southampton, Southampton SO17 1BJ, UK}
\affiliation{STAG  Research  Centre, University  of  Southampton, Southampton SO17 1BJ, UK}
\affiliation{CERN, Theoretical Physics Department, Geneva, Switzerland}

\author{Jonna Koponen}
\affiliation{PRISMA+ Cluster of Excellence \& Institute for Nuclear Physics, Johannes Gutenberg University of Mainz, 55128 Mainz, Germany}

\author{Andreas Kronfeld}
\affiliation{Fermi National Accelerator Laboratory, Batavia, IL 60510, USA}

\author{Christoph Lehner}
\affiliation{Universit\"at Regensburg, Fakult\"at f\"ur Physik, 93040 Regensburg, Germany}

\author{\\Andrew T.~Lytle}
\email{atlytle@illinois.edu}
\affiliation{Department of Physics, University of Illinois at Urbana-Champaign, Urbana, IL 61801, USA}

\author{Guido Martinelli}
\affiliation{University of Roma ``La Sapienza'' and INFN, Sezione di Roma, Piazzale Aldo Moro 5, 00185 Roma, Italy}

\author{Stefan Meinel}
\affiliation{Department of Physics, University of Arizona, Tucson, AZ 85721, USA}

\author{Christopher J.~Monahan}
\affiliation{Department of Physics, William \& Mary, Williamsburg, VA 23187, USA}
\affiliation{Theory Center, Thomas Jefferson National Accelerator Facility, Newport News, VA 23606, USA}

\author{Ethan T.~Neil}
\affiliation{Department of Physics, University of Colorado, Boulder, CO 80309 USA}

\author{\\Antonin Portelli}
\affiliation{Higgs Centre for Theoretical Physics, The University of Edinburgh, EH9 3FD, UK}

\author{James N.~Simone}
\affiliation{Fermi National Accelerator Laboratory, Batavia, IL 60510, USA}

\author{Silvano Simula}
\affiliation{Istituto Nazionale di Fisica Nucleare, Sezione di Roma Tre, Via della Vasca Navale 84, I-00146 Rome, Italy}

\author{Rainer Sommer}
\affiliation{John von Neumann Institute for Computing (NIC),\,DESY,\,Platanenallee 6,\,15738 Zeuthen,\,Germany}
\affiliation{Institut für Physik, Humboldt-Universität zu Berlin, Newtonstr.~15, 12489 Berlin, Germany}

\author{Amarjit Soni}
\affiliation{Physics Department, Brookhaven National Laboratory, Upton, NY 11973, USA}

\author{\\J.~Tobias Tsang}
\affiliation{CP3-Origins and IMADA, University of Southern Denmark, Campusvej 55, 5230 Odense M, Denmark}

\author{Ruth S.~Van de Water}
\affiliation{Fermi National Accelerator Laboratory, Batavia, IL 60510, USA}

\author{Alejandro Vaquero}
\affiliation{Department of Physics and Astronomy, University of Utah, Salt Lake City, UT 84112, USA}

\author{Ludovico Vittorio}
\affiliation{Scuola Normale Superiore, Piazza dei Cavalieri 7, I-56126, Pisa, Italy
 and Istituto Nazionale di Fisica Nucleare, Sezione di Pisa, Largo Bruno Pontecorvo 3, I-56127 Pisa, Italy}

\author{Oliver Witzel}
\email{oliver.witzel@uni-siegen.de}
\affiliation{Center for Particle Physics Siegen, Theoretische Physik 1,
  Universität Siegen,  57068 Siegen, Germany}

\preprint{CERN-TH-2022-036\qquad FERMILAB-CONF-22-433-SCD-T\qquad
JLAB-THY-22-3582} 
\preprint{\qquad MITP-22-020\qquad MIT-CTP/5413 \qquad MS-TP-22-07 \qquad
SI-HEP-2022-11}

\date{\today}
\begin{abstract}
  Lattice quantum chromodynamics has proven to be an indispensable method to
  determine nonperturbative strong contributions to
  weak decay processes. In this white paper for the Snowmass community
  planning process we highlight achievements and future avenues of
  research for lattice
  calculations of weak $b$ and $c$ quark decays, and point out how
  these calculations will help to address the anomalies currently in
  the spotlight of the particle physics community. With future
  increases in computational resources and algorithmic improvements,
  percent level (and below) lattice determinations will play a central
  role in constraining the standard model or identifying new physics.
\end{abstract}
\maketitle

\section{Introduction}
Processes involving weak decays of $b$ or $c$ quarks may provide a
window on new physics not described by the standard model (SM) of
elementary particle physics. For several years such weak decay
processes have shown persistent differences of a few standard
deviations between theoretical predictions of the SM and experimental
measurements. The most prominent deviations, commonly referred to as
$B$ anomalies, include
\begin{itemize}
\item Ratios testing lepton flavor universality for tree-level decays
  such as $B\to D^{(*)} \ell\nu$
\item Tests of lepton flavor universality for rare, loop-level decays
  such as $B \to K^{(*)} \ell^+ \ell^- $
\item Differences in certain $q^2$ bins/ranges for rare decay differential branching fractions\newline
  e.g.~$B\to K^*\ell^+\ell^-$, $B_s\to\phi \ell^+\ell^-$ and corresponding derived
  angular observables like $P_5'$
\item  Some tension in the branching fraction for the rare leptonic
  decay $B_s\to\mu^+\mu^-$
\item Tension between exclusive and inclusive determinations of CKM matrix
  elements $|V_{ub}|$ and  $|V_{cb}|$
\end{itemize}
In addition, the use of QCD factorization to describe nonleptonic decays
is under scrutiny due to observed large discrepancies with experimental results.
Summaries and further details can be found, for example, in
\cite{USQCD:2019hyg,Gambino:2020jvv,Lenz:2021bkv} and in recent
reviews of lattice calculations~\cite{Aoki:2021kgd,Wingate:2021ycr,Lytle:2020tbe,Witzel:2020msp}.
Currently no single quantity is considered significant and trustworthy
enough to claim a smoking gun signal for new physics. While deviations
in ratios testing lepton flavor universality mostly point in the same
direction and collectively favor some SM extensions over others,
the tension between exclusive and inclusive determinations of
Cabbibo-Kobayashi-Maskawa (CKM) matrix elements lacks a good
phenomenological explanation and may hint at underestimated
uncertainties. Understanding and resolving the nature of these $B$
anomalies is the challenge for the coming years.

With ongoing and future experimental measurements from Belle~II, LHCb,
ATLAS, CMS, and BES~III, it is critical for theoretical
predictions to improve to fully leverage increased experimental
precision. A key ingredient here are SM predictions for
contributions due to quantum chromodynamics (QCD), which describes the
strong interactions of quarks and gluons. Standard perturbative methods
work reliably only at (very)
high energies and truly nonperturbative concepts are required to study
the low energy range. Lattice field theory (LFT) is a
nonperturbative framework to study QCD processes at low as well as at
high energies. Based on first principles, LFT uses the QCD Lagrangian
to simulate the strong interaction using Markov
chain Monte Carlo methods. After using a few experimental quantities
to fix input values like bare quark masses, many predictions for QCD
processes can be calculated and the accuracy of the results can be
systematically improved.

Specifically, lattice QCD provides theoretical
input that enables us to determine parameters of the SM such as
the renormalized quark masses, as well as quantities parametrizing
nonperturbative hadronic properties like decay constants, form factors,
bag parameters, or the QCD contribution to lifetimes.
Precise knowledge of such quantities is essential to enhance our
understanding of the SM and distinguish, for example, QCD effects from
new physics. In the remainder of this section, we summarize some of the major
achievements of and opportunities for lattice QCD for weak $b$ and $c$ decays,
and refer the reader to the relevant part of Section~\ref{sec.pac} for
more details.

\medskip
In the heavy quark sector, lattice determinations of the leptonic
decay constants
$f_B$ and $f_{B_s}$ are needed for SM predictions of the rare processes
$B \to \mu^+ \mu^-$ and $B_s \to \mu^+ \mu^-$. Here the lattice community has
managed to determine both decay constants to the sub-percent level
($\sim 0.6 (0.7) \%$ for $f_{B}$($f_{B_s}$)~\cite{Aoki:2021kgd}),
so that hadronic uncertainties are now sub-dominant to other sources
of error. 
For $D_{(s)}$ mesons, $f_D$ and $f_{D_s}$ are used to extract 
$|V_{cd}|$ and $|V_{cs}|$ from leptonic decay measurements. 
Here also the uncertainties in the decay constants are well below those from experiment. 
Further progress can be achieved by
including quantum electrodynamics (QED) and strong-isospin breaking effects into the lattice calculations,
and significant advances have been made in this 
direction~\cite{Carrasco:2015xwa,Giusti:2017dwk,DiCarlo:2019thl,Desiderio:2020oej,Frezzotti:2020bfa,Frezzotti:2021slr,Gagliardi:2022szw}.
The status and physics impact of heavy meson leptonic decays are
expanded on in Section~\ref{sec:ldecays}.

Semileptonic decay processes are critical inputs for heavy flavor studies,
where lattice predictions allow for extraction of CKM matrix elements
and give pure SM predictions of $R$-ratios and other quantities under study.
The most precise exclusive determinations of $|V_{ub}|$ and $|V_{cb}|$
come from combining experimental and lattice results for $B \to \pi \ell \nu$
and $B \to D^{(*)} \ell \nu$ respectively.
In recent years, LHCb has given first measurements of processes such as
$B_c \to J/\psi \ell \nu$, $B_s \to D_s^{(*)} \ell \nu$, $B_s \to K \ell \nu$,
and heavy baryon decays. The lattice community has kept pace with theoretical
calculations of these same processes. Progress and outlook for this important
class of decays is explored in more detail in Section~\ref{sec:sldecays}.

Meson mixing and lifetimes are discussed in~\ref{sec:mixing}.  For neutral $B$-mixing, which
is dominated by short-distance operators, lattice QCD has already delivered ratios of
mass differences with precision around $1.5\%$, compared to experimental
uncertainties of around $0.4\%$. The dominant sources of systematic error in the
lattice QCD calculations can be reduced or eliminated with modern techniques.
The next five years are likely to see high-precision calculations of both bag
parameters ($<1\%$) and for ratios ($<0.5\%$), bringing results to point where QED
effects become important. Neutral $D$-meson mixing presents a greater challenge
both experimentally and theoretically. The short-distance, CP-violating ($\Delta C=2$)
matrix elements have already been determined via lattice QCD with roughly $5\%$
statistical precision, comparable to experimental measurements.

Quark masses --
fundamental parameters of the SM important for high precision tests of the Higgs sector --
have achieved an impressive level of precision
($\lesssim 1.0 (0.5) \%$ for $m_c$ ($m_b$)~\cite{Aoki:2021kgd}),
thanks to long-term efforts from the community.
Here also calculations have reached near to the ``QED wall''
where electromagnetic effects must be accounted for. Details of progress in
this area are given in Section~\ref{sec:masses}.

Moving beyond these ``traditional'' areas, members of the community have continued
to innovate and expand the scope of physics accessible to
lattice computation. Important examples of this relevant to studies of heavy
flavor include first-principles computation of radiative decay
processes (Section~\ref{sec:radiative}),
development and implementation of theoretical machinery to handle multi-hadron
states (discussed in Sections~\ref{sec:sldecays} and~\ref{sec:mixing}),
and exploration of methods to determine inclusive decay rates, which would be
invaluable for resolving inclusive/exclusive discrepancies in determinations
of CKM matrix elements (Section~\ref{sec:inclusive}).
Each of these areas herald a significant advance in our ability
to calculate strong processes from first principles, and in the relevant
subsections we have attempted to provide context on the
remaining challenges and potential timeline to make an impact on phenomenology.

We close by briefly touching on the computational aspects needed to pursue the
outlined calculations in Sec.~\ref{sec.computing}.

\section{Prospects and challenges}\label{sec.pac}
Lattice calculations with charm and bottom quarks face the 
challenge that in order to keep discretization effects in simulations
with fully relativistic actions under control, the quark mass $m_q$
must obey $m_q < a^{-1}$. Here $a^{-1}$ is the inverse lattice
spacing or cutoff typically given in [GeV], whereas the lattice spacing
$a$ is quoted in [fm]. In the past, but also in certain calculations
today, the large mass of charm and especially bottom quarks make it
impossible to meet this requirement, forcing the use of effective
actions. By now algorithmic improvements and increased computational
power enable the use of a fully relativistic setup for all quarks and
more fully relativistic calculations will be published in the near
future. A fully relativistic setup features a simpler and more accurate
handling of the renormalization, which for most calculations will be
performed nonperturbatively. By combining simulations either featuring
up/down quarks at their physical mass or close-to-physical mass bottom
quarks, we can already today largely eliminate two major sources of
uncertainty: chiral extrapolation and the need for (partly)
perturbative renormalization schemes at low energies. By further
decreasing the lattice spacing to $a \le 0.044\,\text{fm}$
($a^{-1} \gtrsim 4.5\, \text{GeV}$), even bottom quarks can be simulated with
the same action as up/down quarks.
With further improved numerical
performance, fully dynamical simulations with up/down, strange, charm,
and bottom quarks become possible~\cite{Chiu:2020tml}, although
  a practical improvement due to simulating dynamical bottom quarks is
most likely marginal. In addition
machine learning techniques may offer new possibilities for LFT
\cite{Boyda:2022nmh}. Complementary to LFT calculations would be
to directly perform quantum simulations~\cite{Bauer:2022hpo}.
That however requires to
have quantum computers with (very) many qubits and long enough
coherence time.

\subsection{Leptonic decays}
\label{sec:ldecays}
Determinations of leptonic decay constants for $D_{(s)}$
\cite{Aubin:2005ar,Follana:2007uv,ETM:2009ptp,Davies:2010ip,ETM:2011zey,FermilabLattice:2011njy,PACS-CS:2011ngu,Na:2012iu,ETM:2013jap,Heitger:2013oaa,FermilabLattice:2014tsy,Carrasco:2014poa,Yang:2014sea,Chen:2014hva,Bazavov:2017lyh,Boyle:2017jwu,Blossier:2018jol,Balasubramamian:2019wgx,Chen:2020qma},
$B_{(s)}$
\cite{Gamiz:2009ku,ETM:2009sed,ETM:2011zey,FermilabLattice:2011njy,McNeile:2011ng,Na:2012kp,ETM:2013jap,ALPHA:2014lwy,Aoki:2014nga,Christ:2014uea,Dowdall:2013tga,ETM:2016nbo,Hughes:2017spc,Bazavov:2017lyh,Balasubramamian:2019wgx},
and $B_c$ \cite{McNeile:2012qf,Colquhoun:2015oha} mesons, obtained from
2-point lattice correlation functions at zero momentum,
showcase the potential of lattice QCD calculations.
Several groups have determined decay constants with high precision
and a complete error budget. The agreement between the different results
strengthens the credibility of lattice results overall and leads to even
more precise average values presented by the Flavor Lattice Averaging Group (FLAG)~\cite{Aoki:2021kgd}.
Using the lattice averages for $f_D$, $f_{D_s}$, and $f_B$, together
with available experimental data for the corresponding leptonic decays,
provides a way of extracting the CKM matrix elements $|V_{cd}|$,
$|V_{cs}|$, and $|V_{ub}|$, respectively. For all three cases lattice
QCD uncertainties are well below those of experiment.

The most precise determinations of $|V_{cd}|$ at present come
from combining experimental measurements of
$D \to \ell \nu_\ell$ with the
lattice determinations of $f_D$.
Until last year, the most precise values of $|V_{cs}|$ similarly came from
$D_s \to \ell \nu_\ell$ and $f_{D_s}$, but new lattice results for
the semileptonic decay $D \to K \ell \nu$~\cite{Chakraborty:2021qav}
are improving on this (see Section~\ref{sec:sldecays}).
The leptonic determination of $|V_{ub}|$ is not competitive with that
from semileptonic decays, but with improvements in the experimental
precision expected from Belle II, it could help to shed light over the
inclusive-exclusive tension in the determination of that parameter.

With precision around or below the percent level, future progress to reduce
uncertainties will require electromagnetic and strong isospin breaking effects
be accounted for. Lattice calculations combining QED and QCD
in the heavy quark sector have already been demonstrated e.g.\ in the case of
charmonium \cite{Hatton:2020qhk} and bottomonium \cite{Hatton:2021dvg}.
Further details on radiative decays, which lift helicity suppression, 
and radiative corrections are presented in Sec.~\ref{Sec.radiative}.

Lattice determinations of neutral $B$ meson decay constants
are also crucial inputs for the study of rare leptonic decays.
These flavor-changing
neutral current (FCNC) processes are highly suppressed in the
SM, and provide important constraints on new physics.
They are largely determined by the same QCD matrix elements as the decay
constants, with corrections from subleading operators. The branching
ratio for $B_s\to\mu^+\mu^-$ is rather precisely
determined~\cite{Bobeth:2013uxa,Beneke:2019slt,Dowdall:2019bea}
using the lattice input for $f_{B_s}$, and shows some tension
with the current experimental
result~\cite{CMS:2019bbr,ATLAS:2018cur,LHCb:2017rmj}. For
$B_d\to\mu^+\mu^-$, the theory error is
larger~\cite{Bobeth:2013uxa,Beneke:2019slt}, but the result is consistent
with the less well-determined experimental
value~\cite{ATLAS:2018cur,LHCb:2017rmj}. Similarly to the extraction
of CKM matrix elements, in these  comparisons lattice QCD inputs have
now exceeded the precision of corresponding experimental measurements.
Further insight on the above theory-experiment tension could
be extracted from a correlated analysis with the parameters that
describe $B_{(s)}$ meson mixing~\cite{Bobeth:2021cxm}.

\subsection{Exclusive semileptonic decays at tree- and loop-level}
\label{sec:sldecays}
Semileptonic decays provide a rich variety of hadronic systems
to study many different decay processes, extract CKM matrix elements,
and perform stringent tests on the SM.
To extract CKM matrix elements, experimental results for tree-level branching
fractions are combined with form factors calculated using lattice QCD.
These combinations often provide the most precise determinations of the relevant
CKM matrix elements, as for
$|V_{cs}^{\text{excl.}}|$, $|V_{ub}^{\text{excl.}}|$, or $|V_{cb}^{\text{excl.}}|$~\cite{Aoki:2021kgd}.
Both tree-level weak charged current and loop-suppressed flavor-changing
neutral current (FCNC) semileptonic decays provide tests of the SM
via comparison of experimental measurements and SM predictions for differential
rates, angular distributions, or ratios of decays with the same hadronic final
state but different generations of final-state leptons.
These ratios test lepton flavor univerality (LFU) and have received substantial
attention due to few-$\sigma$ tensions between experiment and theoretical predictions
for several decay channels.
Several experiments have reported such ratios for tree-level decays
(e.g.~$B\to D^{(*)}\ell\nu$
\cite{Belle:2019gij,LHCb:2015gmp,LHCb:2017rln,BaBar:2013mob} or
$B_c\to J/\psi\ell\nu$ \cite{LHCb:2017vlu}) as well as rare loop-level
$b\to \{s,d\}\ell\ell$ decays (e.g.~$B\to K^{(*)} \ell^+\ell^-$
\cite{LHCb:2014cxe,LHCb:2016ykl,LHCb:2015svh,Belle:2016fev,ATLAS:2017dlm,CMS:2017ivg}
or $B_s\to \phi\ell^+\ell^-$~\cite{LHCb:2015wdu,LHCb:2021zwz})
including also baryonic initial and final states
($\Lambda_b^0\to p K^- \ell^+\ell^-$ \cite{LHCb:2019efc}
or $\Lambda_b^0 \to \Lambda_c^+ \tau^- \bar{\nu}_\tau$~\cite{LHCb:2022piu}).
On the theory side, these ratios are exceptionally clean, and reported
tensions with experimental observations have
increased interest in those quantites.
While tensions vary for different processes, it is intriguing that
these can be accounted for in a model-independent
way by assuming new-physics contributions to certain Wilson coefficients
of the effective weak Hamiltonian. For details see, e.g., Refs.~\cite{Capdevila:2017bsm,Aebischer:2019mlg,Ciuchini:2019usw,Murgui:2019czp,Isidori:2021vtc,Altmannshofer:2021qrr}
as well as references within. Global fits to $b\to s\ell\ell$
and $b\to c\ell\nu$ anomalies provide a basis to build new physics models.
Candidates include, for instance, scenarios with a $Z'$ boson \cite{He:1990pn,He:1991qd,Altmannshofer:2014cfa,Crivellin:2015mga}, leptoquarks \cite{Hiller:2014yaa,Alonso:2015sja,Bauer:2015knc,Fajfer:2015ycq,Barbieri:2015yvd,Bhattacharya:2016mcc,Buttazzo:2017ixm,Crivellin:2017zlb,Angelescu:2018tyl,Cornella:2019hct}, or scenarios related to supersymmetry (SUSY) \cite{Altmannshofer:2017poe,Das:2017kfo,Altmannshofer:2020axr}.
For an overview and further details see Ref.~\cite{Altmannshofer:2022aml}.

To help confirm or refute the observed deviations, higher-precision
calculations of semileptonic form factors are needed, with systematic and
statistical uncertainties commensurate with current and upcoming experiments.
From the perspective of lattice QCD, the simplest processes to compute are
semileptonic decays with a pseudoscalar final state.
These calculations involve two-point and three-point correlation functions at
zero and non-zero momenta, which furnish the two form factors $f_+$ and $f_0$
entering at tree-level or also $f_T$ for rare loop-level decays.
Calculations exist in
the literature for a variety of semileptonic $B$ decays: $B\to\pi\ell\nu$
\cite{Dalgic:2006dt,FermilabLattice:2015mwy,Flynn:2015mha,Colquhoun:2015mfa,Colquhoun:2022atw},
$B\to\pi\ell^+\ell^-$ \cite{FermilabLattice:2015cdh},
$B\to K\ell^+\ell^-$ \cite{Bouchard:2013eph,Bailey:2015dka},
$B_s\to K\ell\nu$
\cite{Bouchard:2014ypa,Flynn:2015mha,Bahr:2016ayy,Monahan:2018lzv,Bahr:2019eom,FermilabLattice:2019ikx},
$B\to D\ell\nu$ \cite{Na:2015kha,MILC:2015uhg,Monahan:2017uby},
$B_s\to D_s\ell\nu$ \cite{Bailey:2012rr,Monahan:2017uby,Monahan:2018lzv,FermilabLattice:2019ikx,McLean:2019qcx,Blossier:2021xvl}
and also for semileptonic
$D$ decays: $D\to \pi\ell\nu$ \cite{FermilabLattice:2004ncd,Na:2011mc,Lubicz:2017syv,Lubicz:2018rfs},
$D\to K\ell\nu$ \cite{FermilabLattice:2004ncd,Na:2010uf,Lubicz:2017syv,Lubicz:2018rfs,Chakraborty:2021qav}.
Once the lattice form factors over the full $q^2$ range have been obtained,
it is a simple post-processing task to integrate these form factors over
the full $q^2$ range to obtain $R$-ratios testing LFU. Hence $R$-ratios have
also been determined for processes like $B\to\pi\ell\nu$ which so far have
not been reported by experiments.
The extraction of $|V_{ub}^{\text{excl.}}|$ from $B\to\pi\ell\nu$, the most precise channel for
that CKM parameter, has commensurate errors coming from experiment and lattice
QCD form factors~\cite{Aoki:2021kgd}.
For $B\to D\ell\nu$ and the extraction of $|V_{cb}^{\text{excl.}}|$, experimental uncertainty
presently exceeds the theoretical error from lattice QCD~\cite{Aoki:2021kgd}.
However, improved theoretical precision will be crucial in both modes in order
to make full use of expected improvements in experimental data from Belle II.
Improved precision will also be valuable for understanding the inclusive-exclusive
tensions for $|V_{ub}|$ and $|V_{cb}|$.
Furthermore, LHCb demonstrated its capabilities to determine the ratio
$|V_{ub}^\text{excl.}/V_{cb}^\text{excl.}|$ by performing a combined analysis of $B_s\to K\mu\nu$
and $B_s\to D_s\mu\nu$ \cite{LHCb:2020ist}. With more statistics and a finer resolution of the $q^2$ bins
this approach can be an interesting alternative to determine the ratio of CKM matrix elements.

In particular, the large mass of the $B_{(s)}$
meson in the initial state leads to a large allowed range of
momentum transfer $q^2$ to the outgoing leptons.
Maintaining statistical control, especially at low $q^2$,
presents a challenge for these calculations. A common approach in the
literature has been to focus on the high-$q^2$
behavior and then extend the calculation to full kinematic range using the
$z$-expansion \cite{Boyd:1994tt,Boyd:1995sq,Caprini:1997mu,Bourrely:2008za}.
Recent work has revived old ideas about using dispersive bounds
\cite{DiCarlo:2021dzg,Martinelli:2021frl,Martinelli:2022tte,Martinelli:2021onb,Martinelli:2021myh} to constrain the low-$q^2$ behavior of the form factors given results at high $q^2$.
Even though covering the full $q^2$ range is computationally challenging,
comparing the shape of the form factors to the experimental data across $q^2$
provides further insight on the quality of our theoretical description of
the experimental process.
Thanks to the advances in simulating heavy flavors and due to
new ensembles with finer lattice spacings, the
range of directly accessible $q^2$ values is increasing.
For more than a decade, the full kinematic range has been accessible to
lattice QCD calculations of $D$ semileptonic decays.
For heavy-to-heavy decays there has been recent progress towards the full $q^2$
range:
$B_s\to D_s \ell\nu$ \cite{McLean:2019qcx} as well as
$B_c\to B_{s,d}$ \cite{Cooper:2020wnj},
$B_c\to D^0\ell\nu$ and $B_c\to D_s\ell^+\ell^-$ \cite{Cooper:2021bkt}.
Near-term progress on extending the $q^2$ range in lattice
QCD calculations of $B$ semileptonic decays (especially $B$-to-light
decays) will be key to improved determinations of CKM matrix elements and more
stringent tests of the SM.

Exclusive semileptonic decays with vector final states are more
challenging and for many years lattice results
for heavy-to-heavy transitions were available
only at zero recoil
($B\to D^*\ell\nu$ \cite{FermilabLattice:2014ysv,Harrison:2017fmw}
and $B_s\to D_s^*\ell\nu$ \cite{Harrison:2017fmw,Blossier:2021xvl}).
Recently, the first lattice calculation of the form factors for $B\to D^{*}\ell\nu$
going beyond zero recoil was performed in Ref.~\cite{FermilabLattice:2021cdg}.
These results gave the first pure-lattice calculation
of the LFU ratio $R(D^*)$.
Two additional and entirely independent determinations of $B\to D^*\ell\nu$ form factors at
non-zero recoil are expected soon~\cite{Barolo:2022jlqcd,Barolo:2022hpqcd}.
Experimentally  $B\to D^*\ell\nu$ is the preferred channel to extract
$|V_{cb}^{\text{excl.}}|$. Hence the lattice form factor data for $B\to D^*\ell\nu$
beyond zero recoil are critical to shed light on the tension between exclusive
and inclusive determinations of $|V_{cb}|$, compare shapes of the form factors, and
test the different methods to constrain the low-$q^2$ range using more precise data
at high $q^2$. Improved knowledge of the $B\to D^*\ell\nu$ form factors will also
benefit the theory prediction of $R(D^*)$, which presently is in tension with the
experimental value \cite{Bifani:2018zmi}.
Recent results for tree-level decays with vector final states, $B_s\to D_s^*\ell\nu$~\cite{Harrison:2021tol}
and $B_c\to J/\psi\ell\nu$~\cite{Harrison:2020gvo}, include all four form factors
and directly cover most of the physically allowed $q^2$ range.
Both modes provide alternative ways to extract $|V_{cb}^{\text{excl.}}|$ and may provide
useful insight into the theory-experiment tensions for $R(D)$ and $R(D^*)$,
especially given expected experimental results from Belle II and LHCb.
One outstanding challenge for the future is including the
final state's decay width as part of the nonperturbative calculation.
For all three processes described above, the vector final-state particle has
a very narrow width and can be treated with chiral perturbation theory
extended to heavy mesons, or taken to be QCD-stable.
However, the target precision dictated by forthcoming experimental data
will eventually require a more rigorous treatment, especially for the decay to
$K^*$.

The general formalism enabing lattice studies of $1\to 2$ hadronic processes,
like $B \to K^*(\to K\pi)\ell^+\ell^-$, has been developed in
Refs.~\cite{Luscher:1986pf,Luscher:1991cf,Lellouch:2000pv,Lin:2001ek,Christ:2005gi,Hansen:2012tf,Briceno:2014uqa,Briceno:2015csa,Agadjanov:2016fbd,Briceno:2021xlc}.
The formalism provides a rigorous non-perturbative relation between finite-volume
Euclidean quantities calculable in lattice QCD and the physical, infinite-volume
$1\to 2$ decay amplitude.
Compared to form factor calculations with single-hadron
final states, $1\to 2$ hadronic processes require conceptually different
calculations and substantially larger computational effort. For a
detailed discussion see Refs.~\cite{USQCD:2019hyg,Gambino:2020jvv}
and references therein. While such calculations have already been
performed in the light sector, e.g., for $K\to\pi\pi$
\cite{Blum:2011ng, RBC:2015gro,Ishizuka:2018qbn,RBC:2020kdj,RBC:2021acc},
decays of $B_{(s)}$ and $D_{(s)}$ mesons are typically more challenging
because the large decaying meson mass makes additional final states
kinematically allowed. The level of difficulty is mainly determined
by the energy of the two-hadron final state, so semi-leptonic
calculations in which the leptons carry away much of the initial
energy are more accessible. In particular, processes such as
$B \to K^*(\to K\pi)\ell^+\ell^-$  \cite{Agadjanov:2016fbd} and
$B \to \rho\,(\to \pi\pi)\ell\nu$ are natural starting points for
multi-hadron heavy-flavor decays. Progress in calculating the
$K\pi\to K\pi$ scattering amplitude, a required input for the weak
decay into this final state, is reported
e.g.\ in Refs.~\cite{Rendon:2018fem,Wilson:2019wfr}.

The kinematics of purely hadronic heavy-flavor decays presents additional challenges.
However, by working with an unphysical setup (e.g.~heavier-than-physical $u/d$ quarks
or lighter-than-physical $b/c$ quarks) the number of kinematically
allowed final states can also be controlled in other channels.
In this way, the methodology used for calculating $K\to\pi\pi$ can be extended
in steps towards $D\to\pi\pi$, for instance.
However, honest calculation of the physical process
eventually requires a formalism that rigorously treats
all important open channels in the decay, including four-particle
states. In this vein, work is ongoing to extend the general $1 \to 2$
formalism to more particles. The approach to study weak three-hadron
decays, including $K\to \pi\pi\pi$, was recently developed
\cite{Muller:2020wjo,Hansen:2021ofl}.

In the future, this work may open the path to lattice calculations of more advanced
phenomenologically interesting processes such as
$\overline {B^0}\to D^+ \{K^-,\pi^-\}$ \cite{Bordone:2020gao,Belle:2021udv},
or the long distance contribution to neutral $D$-meson mixing.
Long-distance contributions (in the form of charm resonances) also occur
in rare loop-level
decays such as $B\to K^{(*)}\ell^+\ell^-$ and $B_s\to\phi\ell^+\ell^-$,
where typically an operator product expansion (OPE) is used to express
matrix elements of nonlocal operators in terms of local-operator
matrix elements. In
Refs.~\cite{Horgan:2013pva,Horgan:2013hoa,Horgan:2015vla} the local
matrix elements have been determined on the lattice to extract the
seven form factors for $B\to K^*\ell^+\ell^-$ and
$B_s\to\phi\ell^+\ell^-$. In this calculation the vector final state
is treated as a stable particle, not accounting for the associated
systematic uncertainties. Since the observed deviations between theory
and experiment for certain $q^2$ bins have persisted for several
years, it is of utmost importance to have well-founded theory
predictions. Once again, the kinematics in the light sector is more
favorable for lattice calculations, and a proper treatment of long-distance
effects in rare kaon decays has been demonstrated
\cite{Christ:2015aha,Christ:2016mmq,Briceno:2019opb,Feng:2020zdc,Christ:2020hwe,Ma:2021azh,Boyle:2022ccj}.
Very first steps towards the direct computation of nonlocal matrix elements
for $B\to K\ell^+\ell^-$ have been taken in Ref.~\cite{Nakayama:2020hhu}.

In analogy to semileptonic decays of mesons, baryons can decay into a
hadronic final state and a lepton pair. While $B$-factories are mostly
run at the $\Upsilon(4s)$ threshold to create $B\bar B$-pairs, the
experiments at the large hadron collider (LHC) measure decays of any
particles originally created from colliding two protons at high
energies. LHCb has reported several measurements of semileptonic
decays of $\Lambda_b$ and $\Lambda_c$
baryons~\cite{LHCB:2015eia,LHCb:2015tgy,LHCb:2017vhq,LHCb:2017yqf,LHCb:2018jna,LHCb:2022piu},
while BES III has also reported semileptonic $\Lambda_c$ decays
\cite{BESIII:2015ysy,BESIII:2016ffj,BESIII:2018mug}.
Belle \cite{Belle:2021crz} and ALICE \cite{ALICE:2021bli} have reported semileptonic $\Xi_c$ decays.
Lattice calculations of baryons tend to suffer from a more severe
signal-to-noise problem compared to those of mesons
\cite{Parisi:1983ae,Lepage:1989hd}.
Nevertheless, lattice calculations of baryonic decays have been performed for
$\Lambda_c \to n\ell\nu$ and $\Lambda_c \to p\ell^+\ell^-$
\cite{Meinel:2017ggx},
$\Lambda_c\to \Lambda \ell\nu$ \cite{Meinel:2016dqj},
$\Lambda_c \to \Lambda^*\ell\nu$ \cite{Meinel:2021mdj,Meinel:2021grq},
$\Lambda_b \to \Lambda_c \tau \bar \nu$ \cite{Datta:2017aue,Detmold:2015aaa},
$\Lambda_b\to p\ell\bar\nu$ \cite{Detmold:2015aaa,Detmold:2013nia},
$\Lambda_b \to \Lambda \ell^+\ell^-$
\cite{Blake:2022vfl,Blake:2019guk,Detmold:2016pkz,Detmold:2012vy},
$\Lambda_b \to \Lambda_c^*\ell\bar\nu$ 
\cite{Meinel:2021rbm,Meinel:2021mdj}, 
$\Lambda_b \to \Lambda^*\ell^+\ell^-$ \cite{Meinel:2020owd},
  and $\Xi_c \to \Xi \ell \nu$ \cite{Zhang:2021oja}.
Integrating the form factors for semileptonic
baryon decays over the allowed range of $q^2$, $R$-ratios testing lepton
flavor universality can be defined and compared to experimental predictions.
Likewise CKM matrix elements can be extracted by combining the form factors
with experimental data. However, for baryonic decays to enter global averages, additional
calculations performed by independent groups are needed \cite{Charles:2004jd,UTfit:2006vpt,Aoki:2021kgd}.

\subsection{Meson mixing and lifetimes}
\label{sec:mixing}
Although a loop-level process, neutral $B_s$- and $B$-meson mixing is the preferred
experimental channel for extracting the CKM matrix
elements $|V_{ts}|$ and $|V_{td}|$. Experiments measure
oscillation frequencies with high precision, and global averages \cite{HFLAV:2019otj}, dominated by the
latest LHCb results~\cite{LHCb:2013fep,LHCb:2016gsk}, show sub-percent level uncertainties.
In the SM and beyond, the hadronic contribution to these processes is governed
by five local, four-fermion ($\Delta B=2$) operators.
The relevant matrix elements are calculable in lattice QCD via two-point and three-point
correlation functions at zero momentum.
The SU(3)-breaking ratio $\xi$~\cite{Bernard:1998dg}, formed using the ratio of
$B_s$- and $B$-meson mixing parameters, is an important input for global CKM
unitarity triangle fits~\cite{Charles:2004jd,UTfit:2006vpt}.
Lattice calculations of $\xi$ have reached percent-level precision~\cite{Gamiz:2009ku,Albertus:2010nm,Bazavov:2012zs,ETM:2013jap,Aoki:2014nga,FermilabLattice:2016ipl,Boyle:2018knm,Dowdall:2019bea},
but further progress is needed to achieve the same level of precision for the
matrix elements (expressed, e.g., as ``bag parameters'') of the individual mixing processes,
presently determined at the few percent level.
The next five years are likely to see high-precision calculations of both bag
parameters ($<1\%$) and for ratios ($<0.5\%$), bringing results to point where QED
effects become important.

At present, tensions exist among the lattice calculations for some $\Delta B=2$ operators.
Calculations by different groups employ different renormalization schemes,
lattice discretizations, and numbers of dynamical quark flavors
\cite{Dalgic:2006gp,Gamiz:2009ku,ETM:2013jap,Aoki:2014nga,FermilabLattice:2016ipl,Dowdall:2019bea}.
Understanding and resolving these tensions is essential for answering the experimental
question of whether or not new physics is present in netural $B$-meson mixing
\cite{King:2019lal,DiLuzio:2019jyq,King:2019rvk}.
As precision improves, higher dimensional operators of the effective weak
Hamiltonian become important, particularly for determination of the
lifetime difference $\Delta \Gamma$, which can provide a complementary test for the SM.
A pioneering study calculated the dimension-7 operators for neutral meson
mixing~\cite{Davies:2019gnp}, and confirmation by an independent calculation is desirable.

Neutral $D$-meson mixing offers complementary constraints on the CKM matrix.
Hadronic contributions to this process enter in two classes:
short-distance, CP-violating ($\Delta C=2$) matrix elements and
long-distance, CP-preserving ($\Delta C=1$) matrix elements.
The $\Delta C=2$ matrix elements have already been determined via lattice QCD
with roughly $5-10\%$ statistical precision~\cite{Carrasco:2014uya,Carrasco:2015pra,Bazavov:2017weg},
comparable to experimental measurements.
Over the next five years, experimental precision is expected to improve by an order of magnitude~\cite{LHCb:2018hne}.
For continued impact, improved lattice calculations are needed on the same timescale.
The long-distance $\Delta C=1$ contributions present a much harder theoretical
problem, but the kinematically simpler case of kaon-mixing has been investigated
\cite{Christ:2012se,Bai:2014cva,Christ:2015pwa}.
Further development of lattice methods for multi-hadron states will be necessary for direct
calculations (see remarks in the previous subsection).
Support for ongoing theoretical and algorithmic work is needed to enable
controlled lattice QCD calculations of the long-distance $\Delta C=1$ matrix
elements on the ten-year timescale.

$B$-meson lifetimes are important targets for lattice QCD.
Besides the $\Delta B=2$ operators appearing in mixing, calculations of hadron lifetimes
also require $\Delta B = 0$ operators.
In particular, lifetime ratios provide valuable tests of expectations from
heavy quark effective theory (HQET) (see \cite{Lenz:2014jha} for a review).
While the ratios
$\tau(B^+)/\tau(B_s)$ and $\tau(B^0)/\tau(B_s)$ are in good agreement
with the HFLAV average \cite{HFLAV:2019otj}, the recent ATLAS
measurement \cite{ATLAS:2020lbz} deviates substantially from recent
measurements by LHCb \cite{LHCb:2019sgv,LHCb:2019nin} and CMS
\cite{CMS:2020efq}. To bolster confidence in the theory predictions,
currently dominated by QCD sum-rule calculations
\cite{King:2019lal,King:2021jsq}, a state-of-the-art lattice
calculation is desirable. Despite early attempts
\cite{Becirevic:2001fy,Becirevic:2000sj,Flynn:2000hx,DiPierro:1999tb,DiPierro:1998ty},
no lattice calculation with a complete systematic error budget exists to date. A
lattice calculation of lifetimes faces the challenge that operators of
different mass dimension mix under renormalization. A breakthrough on
that issue could be made by taking advantage of the gradient flow
\cite{Narayanan:2006rf,Luscher:2009eq,Luscher:2010iy,Luscher:2013cpa}
and the concept of the short-flow-time expansion
\cite{Harlander:2018zpi,Artz:2019bpr,Kim:2021qae,Harlander:2022tgk} to
define a new, nonperturbative renormalization scheme
\cite{Makino:2014taa,Carosso:2018bmz,Suzuki:2020zue,Hasenfratz:2022wll}. A further challenge arises
from quark-line disconnected contributions, which are notoriously hard
to compute with sufficient statistical precision.

\subsection{\texorpdfstring{$b$ and $c$}{b and c} quark masses}
\label{sec:masses}
In addition to providing SM predictions of heavy meson and baryon properties,
lattice QCD simulations are well-suited to the precision determination of
charm and bottom quark masses. These fundamental parameters
are needed to stringently test the Higgs sector of the SM~\cite{Lepage:2014fla},
by comparing Higgs couplings to $b$ and $c$ quarks measured in experiment
with the determinations of quark masses computed via lattice QCD.
The precision computation of quark masses has made good progress in
recent years~\cite{Lytle:2021nre},
with lattice now delivering charm and bottom mass
to a (sub\nobreakdash-)percent level of precision, laying the groundwork for
future experimental tests.
Measurements from HL-LHC will be able to
pin down coupling to bottom at the few-percent level, and first evidence
of coupling to charm may also be
achievable~\cite{CMS:2022cju,Dainese:2019rgk}.
Next generation accelerators could improve these coupling measurements to a level
roughly commensurate with present lattice
determinations~\cite{Lepage:2014fla,Peskin:2013xra}.

There are now several different strategies
for determining quark mass --- among these are
approaches based on moments of current-current
correlators~\cite{Chakraborty:2014aca,Nakayama:2016atf},
the implementation of momentum subtraction schemes on the
lattice~\cite{Lytle:2018evc,ExtendedTwistedMass:2021gbo},
spectroscopy of heavy meson masses and
HQET~\cite{FermilabLattice:2018est,Brambilla:2017hcq},
nonperturbative HQET determinations~\cite{Bernardoni:2013xba}
and computations involving step-scaling in small
volume~\cite{Campos:2018ahf,Heitger:2021apz}.
These methods, though
all relying crucially on lattice simulation, differ substantially
in approach and are hence subject to differing sources of systematic
uncertainty. The good agreement amongst results
from these approaches~\cite{Aoki:2021kgd}
gives confidence in the robustness of the determinations at this
level of precision. Recently, the effects
of adding QED have been quantified,
introducing small (but significant at this level of precision)
shifts to $m_c$~\cite{Hatton:2020qhk} and $m_b/m_c$~\cite{Hatton:2021syc}.
Moving forward, it will be important to hone the efficacy of
existing strategies and also develop new ideas,
while the widespread use of multiple techniques will help
ensure robust error estimates as values continue to improve.

\subsection{Radiative decays and corrections}\label{Sec.radiative}
\label{sec:radiative}
The ability to calculate radiative decay processes from first principles
is an exciting advance that offers opportunities for precision
flavor physics, BSM physics, and hadronic structure.
The development of lattice QCD methods to calculate
radiative decay processes is relatively recent~\cite{Carrasco:2015xwa}.
The general procedure for the lattice calculations has been demonstrated
\cite{Kane:2019jtj,Desiderio:2020oej,Gagliardi:2022szw}.

The determination of CKM matrix elements from leptonic decays (cf.~Sec.~\ref{sec:ldecays})
at $O(\alpha_{\text{em}})$ requires the evaluation of amplitudes
with a real photon~\cite{Carrasco:2015xwa}. Thus, this
technology can directly address radiative corrections in leptonic
decays and advance lattice calculations beyond the ``QED wall''
for these important processes.
First-principles
computations are in progress or planned for the structure-dependent form factors for
$B \to \ell \nu_\ell \gamma$, $D_{(s)} \to \ell \nu_\ell \gamma$ and
$B_{(s)} \to \ell^+ \ell^- \gamma$ and $D_{(s)} \to \ell^+ \ell^- \gamma$,
with a broad photon energy spectrum~\cite{Kane:2021zee}.
For $B$ decay, an enhancement
of the radiative corrections may be expected due to the
nearby $B^*$ resonance~\cite{Becirevic:2009aq}.
Currently only model-dependent predictions of the decay rates are
available in the literature based on QCD factorization and sum rules
\cite{Lunghi:2002ju,Braun:2012kp,Wang:2016qii,Wang:2018wfj,Shen:2018abs,Khodjamirian:2020hob,Beneke:2020fot,Shen:2020hsp,Carvunis:2021jga,Lu:2021ttf,Janowski:2021yvz,Pullin:2021ebn}.
A fully non-factorized, nonperturbative calculation could lead to
improved precision in the determination of the
corresponding CKM matrix elements.

Adding a hard photon in the final state for leptonic decay of a pseudoscalar
meson lifts helicity suppression \cite{Atwood:1994za}, providing sensitivity
to a larger set of operators in the weak effective Hamiltonian.
For example the processes $B^0 \to \ell^+ \ell^- \gamma$ and
$B_s\to \ell^+ \ell^- \gamma$ probe additional operators beyond those of the
corresponding purely leptonic decays,
which can bear on global fits for $b \to s \ell^+ \ell^-$,
and are well-suited for testing LFU
with light leptons~\cite{Kane:2019jtj}.

Radiative processes also give important information on hadron structure.
For large photon energy the process $B\to\ell\nu_\ell \gamma$ is the cleanest
probe of the first inverse moment, $1/\lambda_B$, of the $B$ meson lightcone
distribution amplitude, an important input for QCD factorization predictions
for nonleptonic $B$ decays \cite{Descotes-Genon:2002crx,Wang:2016qii,Beneke:2018wjp}.
Using the upper limit for
$\mathcal B (B^-\to \ell^- \bar\nu_\ell \gamma,\,E_\gamma>1\,\text{GeV})$ from Belle
\cite{Belle:2018jqd} or a lattice form factor calculation can constrain
$\lambda_B$~\cite{Beneke:2011nf}.
A similar calculation in the charm sector would allow to make comparisons with
BES~III results for $D_{(s)}^+ \to e^+ \nu_e\gamma$
\cite{BESIII:2017whk,BESIII:2019pjk}.
An alternate approach, based on recent developments in computing
$x$-dependent hadron structure, may also provide information on the $B$ and $D$
meson distribution amplitudes~\cite{Constantinou:2022yye}.

\subsection{Inclusive decays}
\label{sec:inclusive}
SM predictions for the CKM matrix elements $|V_{ub}|$ and $|V_{cb}|$ have been
computed based on both inclusive
\cite{Gambino:2013rza,Alberti:2014yda,Belle:2021eni,HFLAV:2019otj} and exclusive
\cite{Dalgic:2006dt,FermilabLattice:2014ysv,FermilabLattice:2015mwy,Flynn:2015mha,FermilabLattice:2021cdg,Aoki:2021kgd} decay channels.
The results have exhibited a long-standing tension, with the size of the
tension varying between computations. Compared to their exclusive counterparts,
inclusive semileptonic decays present an additional theoretical challenge for
lattice QCD. The essential difficulty is extracting Minkowski-space spectral
densities from finite-volume Euclidean correlation functions.
However, novel and promising ideas
\cite{Gambino:2020crt,Hashimoto:2017wqo,Hansen:2017mnd,Hansen:2019idp,DeGrand:2022lmc}
may have overcome this theoretical hurdle, paving the way for calculations of
fully inclusive decay rates from lattice QCD simulations.
The new method also opens the door for further applications, such as moments
of the lepton energy and the hadronic invariant mass.
Exploratory numerical studies now exist \cite{Gambino:2020crt,Maechler:2021kax,Gambino:2022dvu},
raising hopes for future work with physical parameters and controlled
systematic uncertainties.
These calculations may play a significant role in resolving the tension
between inclusive and exclusive determinations of CKM matrix elements.
Methods in this entirely new direction in lattice QCD are still in the early
stages of development. It is conceivable, however, that results with
controlled systematics that are sufficiently precise to allow for meaningful
SM tests could become available in the next decade.

\section{Computational Resources}\label{sec.computing}
The calculations outlined in this white paper require post-exascale computational
resources~\cite{Boyle:2022ncb}. A comprehensive research program on weak $b$ and $c$
decays aimed at percent
(or subpercent) precision requires gauge-field ensembles where the wavefunctions
of both heavy and light degrees of freedom are well resolved and can be studied
without distortion. This translates into gauge fields with
both small lattice spacings $a \le 0.044\,\text{fm}$
($a^{-1} \gtrsim 4.5\, \text{GeV}$)
and with large physical volumes $M_\pi\cdot L >4$.
Such simulations require increased lattice sizes. For example, a $256^3 \times 512$
lattice at a lattice spacing $a = 0.040\,\text{fm}$
($a^{-1} \approx 5\,\text{GeV}$)
would allow for simulation of up/down, strange, charm, and bottom quarks
at their physical mass in a $10\,\text{fm}$ box with
$M_\pi\cdot L\approx 7$.
Such an ensemble would provide sufficient physical distance between
hadronic initial and final states to isolate the required matrix elements
in calculations of form factors or meson mixing.
Moreover, such large lattices will allow for new analysis
concepts based e.g.~on masterfields \cite{Francis:2019muy}.
Simulating all quarks at their physical mass is particularly beneficial for systematic
control of calculations like $B\to\rho\ell\nu$, where no effective field theory
is available to guide extrapolations to physical masses.
The large $10\text{ fm}$ box suppresses finite volume
effects, which is especially critical for studying processes with
multiple hadrons.

Today, lattice simulations already tackle lattice sizes of $96^3\times 192$ or
$144^3\times 288$, and research is ongoing to address
algorithmic and computational challenges when simulating finer and
larger lattices. Due to the algorithmic phenomenon of critical slowing
down, development of new alogirithms is likely needed to accelerate sampling
the QCD path integral in hadronic systems with multiple length scales.
On the computational side, harnessing the rapid increase of computational
(GPU) performance is constrained by the stagnating network
performance. Communication-suppressing algorithms
\cite{Luscher:2003qa,Boyle:2022pai} are promising
candidates to meet that challenge and with an anticipated tenfold
performance increase with the next generation of machines, a $256^3
\times 512$ lattice could already become viable.
Professional software support is essential to ensure that
algorithmic advances are leveraged
to their full potential (e.g., by using advances in generating gauge field
configurations in programs that perform measurment tasks).
Post-exascale computing resources, when combined with new and
more precise experimental data, will enable tests of the SM in
the heavy quark sector with unprecedented precision.

\section*{Acknowledgments}

P.A.B.~and A.S.~have been supported by the U.S.~Department of Energy, Office of Science,
Office of Nuclear Physics under the Contract No.~DE-SC-0012704 (BNL).
B.C.~acknowledges support by STFC consolidated grant ST/P000681/1, the Isaac
Newton Trust, and the Leverhulme Trust ECF scheme.
C.T.H.D.~acknowledges support from the UK Science and Technology Facilities Council grant ST/T000945/1.
L.D.D., R.H.~and M.T.H.~are supported by the U.K.~Science and Technology Facility Council
(STFC) grant ST/P000630/1.
This material is based upon work supported by the U.S.~Department of Energy,
Office of Science, Office of High Energy Physics under Award Number DE-SC-0010005 (T.D.~and E.T.N.).
C.D.\ and A.V.~acknowledge support by the National Science Foundation (NSF) under grant number PHY20-13064.
A.X.K.\ and A.T.L.~acknowledge support by the U.S.\ Department of Energy under grant number DE-SC0015655.
F.E.~received funding from the European Research Council (ERC) under the European
Union's Horizon 2020 research and innovation programme under grant agreement No 757646.
Work supported in part by SRA under Grant No. PID2019-106087GB-C21 and by Junta de
Andaluc\'{\i}a under Grants No.~P18-FR-4314 (FEDER), FQM- 101, and A-FQM-467-UGR18 (E.G.).
S.G.\ acknowledges support by the U.S.\ Department of Energy under grant number DE-SC0010120.
M.T.H.\ is supported by UK Research and Innovation Future Leader Fellowship MR/T019956/1.
J.H.~acknowledges support by the Deutsche Forschungsgemeinschaft (DFG)
through the Research Training Group ``GRK 2149: Strong and Weak
Interactions -- from Hadrons to Dark Matter''.
This material is based upon work supported by the U.S. Department of Energy,
Office of Science, Office of Nuclear Physics under grant Contract Numbers
DE-SC0011090 and DE-SC0021006 (W.J.).
J.K.\ acknowledges support by the European Research Council (ERC) under
the European Union's Horizon 2020 research and innovation program
through Grant Agreement No.~771971-SIMDAMA.
This work was supported in part by the U.S. Department of Energy, Office of
Science, under Contract No. DE-AC02-07CH11359 (A.S.K., J.N.S.\ and R.S.V.).
S.M.~is supported by the U.S. Department of Energy, Office of Science, 
Office of High Energy Physics under Award Number DE-SC0009913.
C.J.M.~is supported in part by USDOE grant No.~DE-AC05-06OR23177, under which Jefferson
Science Associates, LLC, manages and operates Jefferson Lab.
The project leading to this application has received funding from the
European Union's Horizon 2020 research and innovation programme under
the Marie Sk{\l}odowska-Curie grant agreement No 894103 (J.T.T.).

\bibliography{Bphysics}
\end{document}